# Elusive Exciton Insulator States in 1T-$HfTe_2$: Exciton softening, and Symmetry Breaking by *Ab Initio* Methods

*Hong Tang*[*1], *Niraj Pangeni*[1], *Daniel D. Rivera*[2], *Adrienn Ruzsinszky*[†1]

[1]*Department of Physics and Engineering Physics, Tulane University, New Orleans, LA 70118 USA*

[2]*Instituto de Física, Universidade de São Paulo, 05315-970 São Paulo, São Paulo, Brasil*

## ABSTRACT

Recent experiments have provided evidence for excitonic insulator (EI) states in 1T-$HfTe_2$. In this work, we investigate EI states in monolayer, bilayer, trilayer, and bulk 1T-$HfTe_2$ using advanced meta-generalized gradient approximation (meta-GGA) calculations and a model Bethe–Salpeter equation (BSE) approach, together with structural and electronic symmetry-breaking analyses. Our results show that both the monolayer and bilayer exhibit negative exciton energies, leading to the spontaneous formation of bound excitons and EI states, whereas the trilayer and bulk display positive exciton energies and do not support EI states. Structural symmetry-breaking calculations show very small in-plane displacements of the Hf atoms from their symmetric positions in the monolayer and multilayers, consistent with experimental observations. Interestingly, electronic symmetry-breaking calculations for the monolayer, performed using a symmetric structure and a hybrid functional, show a pronounced unfolded valence-band feature at the M point and no unfolded conduction-band states near the Fermi level at Γ, in good agreement with experimental results. Overall, our findings support the existence of EI states in low-dimensional 1T-$HfTe_2$. The methodology developed here can be readily extended to investigate EI behavior in other related quantum material systems.

Two-dimensional (2D) layered transition metal dichalcogenides (TMDs) are a fascinating class of materials exhibiting many novel properties [1–6]. A single layer of 2D TMDs can crystallize in several structural phases, including the trigonal prismatic (1H) phase (such as in $MoS_2$ [7] and $WS_2$ [8]), the octahedral (1T, or anti–trigonal prismatic) phase (e.g., $TiSe_2$ [9] and $HfTe_2$ [6]), and the distorted 1T (1T′) phase [3] (e.g., $WTe_2$ and $MoTe_2$), which arises from lattice distortions induced by metal–metal interactions.

[*] Email: htang5@tulane.edu
[†] Email: aruzsin@tulane.edu

Depending on the intralayer structures and interlayer stacking in the bulk, additional phases can form, including the 2H and 3R phases [10] derived from the 1H structure, as well as the hexagonal 1T [6,9], monoclinic 1T′ [3,11], and orthorhombic Td [11] phases. These diverse intralayer structures and interlayer stacking configurations give rise to a variety of crystal symmetries, resulting in a wide range of physical properties, including Weyl semimetal behavior [11], charge-density-wave (CDW) order [12], the quantum spin Hall effect [3], superconductivity [13], excitonic insulating behavior [6,9], and topological exciton insulators [14]. As a result, 2D TMDs exhibit great potential for novel nanoelectronic and optoelectronic applications.

1T-$HfTe_2$ is one such interesting 2D TMD. Bulk 1T-$HfTe_2$, synthesized using both the self-flux method [15] and the chemical vapor transport method [16], is a semimetal with both hole and electron pockets in the first Brillouin zone (BZ), as revealed by quantum oscillation measurements, angle-resolved photoemission spectroscopy (ARPES) [17], and transport studies [18].

Interesting band quantization induced by potassium intercalation has been visualized by ARPES and is associated with intercalation-induced monolayer and multilayer complex surface structures [17,19]. Monolayer 1T-$HfTe_2$, synthesized by molecular beam epitaxy on different substrates (i.e., InAs, graphene, and $MoS_2$), exhibits a Dirac-cone-like valence-band dispersion within an electron binding energy range of 0.0–1.0 eV. In addition, electron doping from the InAs and graphene substrates shifts the valence bands downward in energy [20].

Bulk 1T-$HfTe_2$ also exhibits a large magnetoresistance at low temperatures, which has been attributed to the compensation between electron and hole carriers as well as the mixed orbital character of the hole states [21].

Possible excitonic insulator (EI) states in monolayer and bilayer 1T-$HfTe_2$ have been observed in low-temperature angle-resolved photoemission spectroscopy (ARPES) measurements [6]. These measurements reveal a band-gap opening around the Γ point and charge-density-wave (CDW)–related unfolded band features around the M point in the Brillouin zone (BZ). At the same time, no structural changes are detected by temperature-dependent Raman spectroscopy and synchrotron X-ray measurements.

Furthermore, pronounced layer-dependent effects have been reported [6], with trilayer and thicker multilayer samples showing no EI states at low temperatures. These experimental results indicate that low-dimensional 1T-$HfTe_2$ hosts an EI phase in the absence of obvious structural or phonon-related contributions to the phase transition.

Direct computational characterization of EI states in 1T-$HfTe_2$ remains scarce. In this work, we perform extensive computational investigations of the EI states in 1T-$HfTe_2$ using density functional theory (DFT) and many-body perturbation theory. Our results confirm the presence of EI states in monolayer and bilayer 1T-$HfTe_2$, and the absence of EI states in trilayer and bulk systems, in agreement with experimental observations. In addition, the role of structure-related effects, particularly structural symmetry breaking associated with the phase transition, is also discussed.

## Results and discussions

Figure 1a presents the calculated band structures along the Γ–M path superimposed on the corresponding experimental ARPES spectra for monolayer, bilayer, and trilayer 1T-$HfTe_2$, showing overall good agreement. Details of the computational methods, as well as the calculated band structures for multilayer and bulk systems, are provided in the Supporting Information (SI), Note and Figures S1–S4.

Notably, the band structures near the Fermi level calculated using the regularized-restored strongly constrained and appropriately normed ($r^2$SCAN) functional [22] including spin–orbit coupling (SOC) show significantly better agreement with the ARPES results than those obtained with the Perdew–Burke–Ernzerhof (PBE) functional [23] with SOC [6,17]. The PBE calculations exhibit substantial overlap between the conduction and valence bands around the Γ and M points, in contrast to experiment. The improved performance of $r^2$SCAN can be attributed to its more accurate treatment of self-interaction effects in the exchange–correlation functional [24], which leads to better agreement with the measured electronic structure.

Based on the $r^2$SCAN+SOC wavefunctions, we calculate exciton energies and exciton binding energies using the robust meta-GGA+mBSE approach. In this method, a model Bethe–Salpeter equation (BSE) is employed, in which the screened Coulomb interaction is obtained from a modeled inverse dielectric function $\varepsilon^{-1}(q)$ as a function of the wavevector $q$, $\varepsilon^{-1}(q) = 1 - (1-\alpha)exp\left(\frac{-q^2}{4\mu^2}\right)$, where $\alpha = \varepsilon^{-1}(q=0)$, and the parameters $\alpha$ and $\mu$ are determined by fitting to random-phase approximation (RPA) calculations. The dielectric function $\varepsilon^{-1}(q)$ is calculated using $r^2$SCAN+SOC; the corresponding fitting parameters are provided in Figures S6–S9 of the SI. Further details of the meta-GGA+mBSE and mBSE methodologies can be found in Refs. [25–27].

The calculated binding energies of the first exciton with momentum transfer from Γ to M are used to assess possible exciton softening. The results are summarized in Table 1. For both monolayer and bilayer 1T-$HfTe_2$, the energy of the first exciton is negative, indicating that the exciton binding energy $E_b$ exceeds the band gap. This condition implies the formation of excitonic insulator (EI) states in the monolayer and bilayer systems. In contrast, trilayer and bulk 1T-$HfTe_2$ exhibit positive first-exciton energies, corresponding to unbound excitons and effectively negative binding energies.

The screening parameter $\alpha$ characterizes the strength of Coulomb screening, with smaller values of $\alpha$ indicating stronger screening effects. As shown in Table 1, the screening strength increases gradually from the monolayer to the bulk. These calculated results are consistent with experimental observations [6], which report a gap opening and CDW-like unfolded valence-band features at the M point for monolayer and bilayer 1T-$HfTe_2$ at low temperatures, but not for trilayer or bulk samples.

In the above calculations, the parameters $\alpha$ and $\mu$ are determined by fitting to the inverse dielectric function $\varepsilon^{-1}(q)$ obtained within the random-phase approximation (RPA), calculated using r²SCAN+SOC. The function $\varepsilon^{-1}(q)$ is evaluated for a supercell geometry, which necessarily includes a large vacuum region for two-dimensional layered systems.

Ghosh *et al.* [28] proposed a scheme to rescale the supercell dielectric tensor $\varepsilon_{\infty}^{SC}$ to an effective two-dimensional form $\varepsilon_{\infty}^{2D}$, given by $\varepsilon_{\infty,\parallel}^{2D} = 1 + \frac{c}{t}\left(\varepsilon_{\infty,\parallel}^{SC} - 1\right)$, and $\varepsilon_{\infty,\perp}^{2D} = \left[1 + \frac{c}{t}\left(\frac{1}{\varepsilon_{\infty,\perp}^{SC}} - 1\right)\right]^{-1}$ where the subscripts $\parallel$ and $\perp$ denote the in-plane and out-of-plane components, respectively, and $\varepsilon_{\infty}$ refers to the ion-clamped optical dielectric constant associated with long-wavelength screening. Here, $c$ and $t$ represent the out-of-plane thicknesses of the supercell and the two-dimensional layer, respectively.

The effective two-dimensional dielectric constant is then defined as $\varepsilon_{\infty}^{2D,eff} = \frac{2\varepsilon_{\infty,\parallel}^{2D}\varepsilon_{\infty,\perp}^{2D}}{\varepsilon_{\infty,\parallel}^{2D} + \varepsilon_{\infty,\perp}^{2D}}$. Within this framework, the screening parameter $\alpha$ is given by $\alpha = 1/\varepsilon_{\infty}^{2D,eff}$. The parameter $\mu$ is calculated as $\mu = aN_e^{1/3}\left[1 - \left(\varepsilon_{\infty}^{2D,eff}\right)^{-1}\right]^{-1/2}$ where $N_e$ is the number of valence electrons of the transition-metal atom (equal to 4 for Hf), and $a$ is an optimized constant taken to be 0.3 [28]. Using the rescaled values of $\alpha$ and $\mu$, we recalculate the exciton energies within the r²SCAN+SOC+mBSE framework. The resulting values are summarized in Table 2.

For the monolayer, the experimental thickness measured by scanning tunneling spectroscopy (STS) is approximately 6.79 Å [29]. We adopt the average interlayer thickness of bulk 1T-$HfTe_2$, 6.779 Å, which is very close to the STS value. For free-standing monolayers or multilayers, the electron density can extend over both surfaces, leading to some uncertainty in defining an effective thickness. To assess the impact of this uncertainty, we consider thickness values ranging from 1.0 to 1.1 times 6.779 Å.

As shown in Table 2, the first exciton energies remain negative within this thickness range, indicating spontaneous formation of bound excitons in the monolayer and the persistence of the excitonic-insulator (EI) state, consistent with the results in Table 1. For the trilayer, the rescaled results also agree with the previous calculations. For the bilayer, the excitonic behavior depends on the assumed thickness; however, within the physically reasonable range considered here, the bilayer can still support an EI state. Overall, the calculations employing the modified dielectric-rescaling scheme remain in good agreement with experimental observations.

The phonon dispersions of monolayer and bilayer 1T-$HfTe_2$ are shown in Figures 1b and 1c (see Figure S5 in the SI for the trilayer). All cases exhibit a dip at the M point, but no imaginary phonon frequencies are present at M, indicating structural stability. This is consistent with experimental observations, which show no detectable structural changes.

The softening modes observed in electron energy loss spectra (EELS) are also closely related to the excitonic insulator (EI) phase transition [9]. We performed $G_0W_0$+BSE calculations based on PBE+U+SOC (with $U = 3.2$ eV) for monolayer 1T-$HfTe_2$. A scissor correction was applied to the indirect band gap between Γ and M to account for the experimentally observed low-temperature gap, which is approximately 80 meV according to Ref. [6].

Electron doping induces a Fermi-level shift of approximately 300 meV [20], thereby introducing significant uncertainty in the experimentally measured gap. We therefore estimate an upper bound of 400 meV for the band gap. Under these conditions, the calculated GW band structure and the lowest exciton band are shown in Figure 1d. A small portion of the exciton band near the M point lies below zero energy, indicating negative exciton energies, i.e., exciton binding energies that exceed the band gap, which is characteristic of an EI state.

The calculated EELS spectra (Figure 1e) exhibit a softening-like behavior in the onset of the energy-loss peaks as the exciton momentum approaches $Q_{\Gamma M}$. The appearance of zero-energy loss peaks in the EELS curves at momenta close to $Q_{\Gamma M}$signals the presence of zero-energy excitation modes, providing further evidence for the EI state.

As currently debated for several excitonic-insulator (EI) candidates, such as 1T-$TiSe_2$ [9,30] and $Ta_2NiSe_5$ [31–33], the structural change observed at low temperatures—detected experimentally and associated with electron–phonon interactions—may play a role in the phase transition. In contrast, low-temperature Raman spectroscopy of 1T-$HfTe_2$ does not show phonon mode changes, thereby presumably excluding structurally driven interactions in the transition.

The Goldstone mode of an EI state corresponds to the excitonic sound mode, which arises from oscillations of the relative phase between bound electrons and holes [34,35]. At temperatures above the EI phase transition, this mode manifests as electronic and plasmon-like excitations and can hybridize with low-energy phonons. Such mixing is weak at zero wavevector but can become significant at the charge-density-wave (CDW) wavevector [9].

In this work, we also investigate the effects of structural changes (perturbed atomic positions) and electronic changes (valence and conduction band occupations) in 1T-$HfTe_2$, as well as their possible interplay, using the symmetry-breaking (SB) method [36].

For structural symmetry breaking (SSB), the calculations start by randomizing the atomic positions, followed by structural relaxation and a self-consistent $r^2$SCAN calculation for a 2×2×1 supercell of monolayer 1T-$HfTe_2$. The resulting band structure is then unfolded back to the 1×1×1 cell.

For electronic symmetry breaking (ESB), a 2×2×1 supercell is constructed from the pristine symmetric 1×1×1 monolayer without any randomization of atomic positions. Electrons from the highest valence band are promoted to the lowest conduction band to create electron–hole pairs, and a hybrid r2SCAN–Y functional (where Y denotes the fraction of exact Hartree–Fock exchange) is employed in the self-consistent calculation of the 2×2×1 supercell. The value of Y is taken to be 28.6%, corresponding to the screening parameter α for the monolayer, as shown in Table 1, since α represents the screening of the long-range, nonlocal exchange interaction [26, 37, 38].

All other computational procedures are the same as those used for the SSB calculations. By ESB, we mean that the symmetric electron charge density is perturbed by electron promotion, leading to a change or breaking of the associated density symmetry.

The hybrid $r^2$SCAN-Y functional can, to some extent, account for electron–hole interactions, since the missing electron–electron interactions associated with the removed electrons (i.e., the holes) in the highest valence band can effectively reflect and represent electron–hole coupling.

For the combined structural and electronic symmetry breaking (SESB) calculations, both perturbed atomic positions and electron promotions are applied, and the hybrid $r^2$SCAN-Y functional is used. Further computational details are provided in the Supporting Information (SI).

The calculated unfolded band structures are shown in Figure 2. For SSB (Figure 2a), the unfolded band structure closely resembles that of the primitive 1×1×1 monolayer cell, with only faint unfolded valence-band features appearing at the M point. The maximum in-plane deviation of the Hf atoms from their symmetric positions in this monolayer is 0.003 Å, a value that is likely below the detection limit of synchrotron X-ray measurements [39]. The relative spectral weight of the unfolded valence-band features at M ($W_M$, see Figure 2a) is only 0.27%, compared with approximately 100% for the conduction band at M at slightly higher energy. This indicates that the SSB effect alone is unlikely to account for the relatively strong CDW-like unfolded valence-band feature observed experimentally at M.

In addition, a faint unfolded conduction-band feature appears at Γ in Figure 2a. This feature should manifest as an electron pocket; however, no such electron pocket at Γ is observed experimentally, in contrast to the electron pocket at M reported in Figure 2 of Ref. 6.

For ESB (Figure 2b), the calculated unfolded valence-band structure again closely resembles that of the primitive 1×1×1 monolayer cell, whereas the conduction bands exhibit pronounced modifications. Clear unfolded valence-band features appear at M, while no apparent unfolded conduction-band features are found at Γ or near the Fermi level. The relative spectral weight of the unfolded valence-band features at M is 33.2% ($W_M$, see Figure 2b), which is substantial and consistent with the experimental observations.

For the combined SESB case (Figure 2c), the unfolded valence-band structure also remains similar to that of the primitive 1×1×1 monolayer cell, while the conduction bands again show strong variations. More complex features emerge around the Fermi level and the M point, arising from the combined effects of electron promotion and atomic position perturbations. The relative spectral weight of the unfolded valence-band features at M is reduced to 6.9%, which may nevertheless remain experimentally detectable.

Overall, this analysis indicates that the ESB effect is dominant, pointing toward an excitonic-insulator state driven by strong electron–hole interactions in monolayer 1T-$HfTe_2$.

The SSB calculation for bilayer 1T-$HfTe_2$ is shown in Figure 2d, where very faint unfolded valence-band features appear around the M point. The maximum in-plane deviation of the Hf atoms from their symmetric positions is only 0.002 Å, which is extremely small and may not be experimentally detectable. This result is therefore consistent with experimental observations.

The SSB calculation for trilayer 1T $HfTe_2$ is presented in Figure 2e; similarly, it shows very faint unfolded valence-band features around M, accompanied by a maximum in-plane deviation of the Hf atoms of only 0.002 Å. These results demonstrate consistency between the SSB calculations and experimental observations for both bilayer and trilayer 1T $HfTe_2$, confirming that structural changes play a relatively minor role in this system.

In conclusion, we investigated excitonic-insulator (EI) states in 1T-$HfTe_2$ in its monolayer, bilayer, trilayer, and bulk forms using advanced meta-GGA $r^2$SCAN calculations, model Bethe–Salpeter equation (BSE) approaches, and structural and electronic symmetry-breaking schemes. Our results show that both the monolayer and bilayer exhibit negative exciton energies for the lowest-energy excitons, indicating exciton binding energies larger than the band gap and supporting the existence of EI states. In contrast, the trilayer and bulk systems display positive exciton energies, indicating the absence of EI states.

Structural symmetry-breaking calculations for the monolayer, bilayer, and trilayer reveal minute in-plane deviations of the Hf atoms from their symmetric positions, consistent with experimental observations. In contrast, the electronic symmetry-breaking calculation for the monolayer with a symmetric structure, performed using the hybrid $r^2$SCAN-Y functional, reproduces the strong unfolded valence-band feature at the M point while showing no unfolded conduction-band features at Γ or near the Fermi level, in good agreement with experiment.

Overall, our results support the presence of excitonic-insulator states in low-dimensional 1T-$HfTe_2$. The methodology employed here is general and can be applied to the analysis of other related material systems.

**Acknowledgements**

For the work on excitonic insulators, A.R. acknowledges the U.S. Department of Energy, Office of Science, Office of Basic Energy Sciences, under Award Number DE-SC0026293. The work of H.T. was supported by the Department of Energy Office of Basic Sciences under grant no. DE-SC0018331. The computational work is supported by the resources at the National Energy Research Scientific Computing Center (NERSC) high performance computers (HPC) and the Louisiana Optical Network Infrastructure (LONI) HPC. D. D. R. acknowledges support from agencies FAPESP (processes 2023/03493-0, 2025/08647-0 and 2023/09820-2) and CNPq.

**References**

1. A. Chaves, J. G. Azadani, H. Alsalman, D. R. da Costa, R. Frisenda, A. J. Chaves, S. H. Song, Y. D. Kim, D. He, J. Zhou, A. Castellanos-Gomez, F. M. Peeters, Z. Liu, C. L. Hinkle, S.-H. Oh, P. D. Ye, S. J. Koester, Y. H. Lee, P. Avouris, X. Wang and T. Low, Bandgap Engineering of Two-Dimensional Semiconductor Materials, npj 2D Materials and Applications, 2020, 4, 29.
2. G. Wang, A. Chernikov, M. M. Glazov, T. F. Heinz, X. Marie, T. Amand and B. Urbaszek, Colloquium: Excitons in Atomically Thin Transition Metal Dichalcogenides, *Rev. Mod. Phys.,* 2018, **90**, 021001.
3. X. Qian, J. Liu, L. Fu, and J. Li, Quantum spin Hall effect in two-dimensional transition metal dichalcogenides, Science (2014) 346, 6215, 1344-1347.
4. X.-X. Zhang, T. Cao, Z. Lu, Y.-C. Lin, F. Zhang, Y. Wang, Z. Li, J. C. Hone, J. A. Robinson, D. Smirnov, S. G. Louie and T. F. Heinz, Magnetic Brightening and Control of Dark Excitons in Monolayer WSe2, Nature Nanotech., 2017, 12, 883–888.
5. C. Jin, Z. Tao, T. Li, Y. Xu, Y. Tang, J. Zhu, S. Liu, K. Watanabe, T. Taniguchi, J. C. Hone, L. Fu, J. Shan and K. F. Mak, Stripe Phases in WSe2/WS2 Moiré Superlattices, Nature Materials, 2021, 20, 940–944.
6. Q. Gao, Y. Chan, P. Jiao, H. Chen, S. Yin, K. Tangprapha, Y. Yang, X. Li, Z. Liu, D. Shen, S. Jiang, and P. Chen, Observation of possible excitonic charge density waves and metal–insulator transitions in atomically thin semimetals, Nature Physics, 20, (2024) 597-602.
7. Mak, K. F., Lee, C., Hone, J., Shan, J. & Heinz, T. F. Atomically Thin MoS2: A New Direct-Gap Semiconductor. Phys. Rev. Lett. 105, 136805 (2010).
8. Kim, H. C. et al. Engineering optical and electronic properties of WS2 by varying the number of layers. ACS Nano 9, 6854–6860 (2015).
9. A. Kogar, et al. Signatures of exciton condensation in a transition metal dichalcogenide. Science 358,1314–1317 (2017).
10. Z. H. Peng, M. Cotrufo, D. Xu, S. A. Mann, S. Qiu, D. N. Basov, M. Delor, A. Alú, P. J. Schuck, and C. Trovatello, 3R-stacked transition metal dichalcogenide non-local metasurface for efficient second-harmonic generation, Nature Photonics 19, 1376–1384 (2025).
11. E. J. Sie, C. M. Nyby, C. D. Pemmaraju, S. J. Park, X. Shen, J. Yang, M. C. Hoffmann, B. K. Ofori-Okai, R. Li, A. H. Reid, S. Weathersby, E. Mannebach, N. Finney, D. Rhodes, D. Chenet, A. Antony, L. Balicas, J. Hone, T. P. Devereaux, T. F. Heinz, X. Wang, A. M. Lindenberg, An ultrafast symmetry switch in a Weyl semimetal, Nature 565, 61–66 (2019).
12. J. Hwang, W. Ruan, Y. Chen, S. Tang, M. F Crommie, Z.-X. Shen, and S.-K. Mo, Charge density waves in two-dimensional transition metal dichalcogenides, 2024 Rep. Prog. Phys. 87 044502.
13. S. Simon, H. Yerzhakov, K. P. Sajilesh, A. Vakahi, S. Remennik, J. Ruhman, M. Khodas, O. Millo, H. Steinberg, The transition-metal-dichalcogenide family as a superconductor tuned by charge density wave strength, Nature Communications volume 15, Article number: 10439 (2024).

14. D. Varsano, M. Palummo, E. Molinari, and M. Rontani, A monolayer transition-metal dichalcogenide as a topological excitonic insulator, Nature Nanotechnology, 15, 367–372 (2020).
15. Q. Li, G. Jin, N. Tang, B. Wang, B. Shen, D. Guo, D. Zhong, H. Zuo, and H. Wang, Shubnikov–de Haas oscillations and planar Hall effect in HfTe2, Phys. Rev. B 108, 235155 (2023).
16. K. Ueno, Introduction to the Growth of Bulk Single Crystals of Two-Dimensional Transition-Metal Dichalcogenides, J. Phys. Soc. Jpn. 84, 121015 (2015).
17. Y. Nakata, K. Sugawara, A. Chainani, K. Yamauchi, K. Nakayama, S. Souma, P.-Y. Chuang, C.-M. Cheng, T. Oguchi, K. Ueno, T. Takahashi, and T. Sato, Dimensionality reduction and band quantization induced by potassium intercalation in 1T-HfTe2, Phys. Rev. Materials 3, 071001(R) (2019).
18. P. C. Klipstein, D. R. P. Guy, E. A. Marseglia, J. I. Meakin, R. H. Friend, and A. D. Yoffe, Electronic properties of HfTe2, J. Phys. C: Solid State Phys. 19 (1986) 4953-4963.
19. Z. E. Youbi, S. W. Jung, S. Mukherjee, M. Fanciulli, J. Schusser, O. Heckmann, C. Richter, J. Minár, K. Hricovini, M. D. Watson, and C. Cacho, Bulk and surface electronic states in the dosed semimetallic HfTe2, Phys. Rev. B 101, 235431 (2020).
20. P. Tsipas, P. Pappas, E. Symeonidou, S. Fragkos, C. Zacharaki, E. Xenogiannopoulou, N. Siannas, and A. Dimoulas, Epitaxial HfTe2 Dirac semimetal in the 2D limit, APL Mater. 9, 101103 (2021).
21. S. Mangelsen, P. G. Naumov, O. I. Barkalov, S. A. Medvedev, W. Schnelle, M. Bobnar, S. Mankovsky, S. Polesya, C. Näther, H. Ebert, and W. Bensch, Large nonsaturating magnetoresistance and pressure-induced phase transition in the layered semimetal HfTe2, Phys. Rev. B 96, 205148 (2017).
22. Furness, J. W.; Kaplan, A. D.; Ning, J.; Perdew, J. P.; Sun, J. Accurate and Numerically Efficient r2SCAN Meta-Generalized Gradient Approximation, J. Phys. Chem. Lett. **2020**, 11, 8208−8215.
23. Perdew, J. P.; Burke, K.; Ernzerhof, M. Generalized Gradient Approximation Made Simple, Phys. Rev. Lett. **1996**, 77, 3865.
24. M. Kothakonda, A. D. Kaplan, E. B. Isaacs, C. J. Bartel, J. W. Furness, J. Ning, C. Wolverton, J. P. Perdew, J. Sun, Testing the r2SCAN Density Functional for the Thermodynamic Stability of Solids with and without a van der Waals Correction, ACS Mater. Au 2023, 3, 102–111.
25. A. Tal, P. Liu, G. Kresse, and A. Pasquarello, Accurate optical spectra through time-dependent density functional theory based on screening-dependent hybrid functionals, Phys. Rev. Research 2, 032019(R) (2020).
26. W. Chen, G. Miceli, G.M. Rignanese, and A. Pasquarello, Nonempirical dielectric-dependent hybrid functional with range separation for semiconductors and insulators, Phys. Rev. Mater. 2, 073803 (2018).
27. Tang, H.; Yin, L.; Csonka, G. I.; Ruzsinszky, A. Exploring the exciton insulator state in 1T-TiSe2 monolayer with advanced electronic structure methods, Phys. Rev. B 2025, 111, L201401.

28. A. Ghosh, S. Jana, M. Hossain, D. Rani, S. Śmiga, and P. Samal, Advancing excited-state properties of two-dimensional materials using a dielectric-dependent hybrid functional, Phys. Rev. B 112, 045128, 2025.
29. S. Aminalragia-Giamini, J. Marquez-Velasco, P. Tsipas, D. Tsoutsou, G. Renaud, and A. Dimoulas, Molecular beam epitaxy of thin HfTe2 semimetal films, 2017 2D Mater. 4 015001.
30. D. Pashov, R. E. Larsen, M. D. Watson, S. Acharya, and M. van Schilfgaarde, TiSe2 is a band insulator created by lattice fluctuations, not an excitonic insulator, npj Computational Materials volume 11, Article number: 152 (2025).
31. L. Windgätter, M. Rösner, G. Mazza, H. Hübener, A. Georges, A. J. Millis, S. Latini, and A. Rubio, Common microscopic origin of the phase transitions in Ta2NiS5 and the excitonic insulator candidate Ta2NiSe5, npj Computational Materials volume 7, Article number: 210 (2021).
32. E. Baldini, A. Zong, D. Choi, C. Lee, M. H. Michael, L. Windgaetter, I. I. Mazin, S. Latini, D. Azoury, B. Lv, A. Kogar, Y. Su, Y. Wang, Y. Lu, T. Takayama, H. Takagi, A. J. Millis, A. Rubio, E. Demler, and N. Gedik, The spontaneous symmetry breaking in Ta2NiSe5 is structural in nature, PNAS 2023 Vol. 120 No. 17 e2221688120.
33. Y. F. Lu, H. Kono, T. I. Larkin, A. W. Rost, T. Takayama, A. V. Boris, B. Keimer & H. Takagi, Zero-gap semiconductor to excitonic insulator transition in Ta2NiSe5, Nature Communications volume 8, Article number: 14408 (2017).
34. J. Goldstone, A. Salam, and S. Weinberg, Broken Symmetries, Phys. Rev. 127, 965, 1962.
35. B. Remez and N. R. Cooper, Effects of disorder on the transport of collective modes in an excitonic condensate, Phys. Rev. B 101, 235129, 2020.
36. J. P. Perdew, SCAN meta-GGA, strong correlation, symmetry breaking, self-interaction correction, and semi-classical limit in density functional theory: Hidden connections and beneficial synergies? APL Comput. Phys. 1, 010903 (2025).
37. H. R. Gopidi, R. Zhang, Y. Wang, A. Patra, J. Sun, A. Ruzsinszky, J. P. Perdew, and P. Canepa, Reducing self-interaction error in transition-metal oxides with different exact-exchange fractions for energy and density, Phys. Rev. B 113, 165115 (2026).
38. H. Tang, N. Pangeni, and A. Ruzsinszky, Meta-Generalized Gradient approximations with a Dielectric Function Model in the Time-Dependent Density Functional Theory for Optical Properties of Solid Materials, J. Phys. Chem. C 2025, 129, 17296−17305.
39. S. Lai, X. Yang, J. Shi, S. Liu, Y. Guo, L. Yan, J. Zang, Z. Zhang, Q. Jia, J. Sun, S. Cheng, and C. Shan, Bulk hexagonal diamond, Nature 651, 621-625 (2026)

Tables

Table 1. The screening parameters $\alpha$ and $\mu$ and the calculated band gap $E_g$, the first exciton energy $E_{exc}$, the binding energy of the first exciton $E_b$ ($E_b = E_g - E_{exc}$), and the EI states of monolayer, bilayer, tri-layer and bulk 1T-$HfTe_2$ with the $r^2$SCAN+SOC and mBSE methods.

| | Monolayer | Bilayer | Tri-layer | Bulk |
|---|---|---|---|---|
| $\alpha$ | 0.286 | 0.215 | 0.149 | 0.021 |
| $\mu$ (1/Å) | 0.350 | 0.350 | 0.450 | 0.980 |
| $E_g$ (eV) | 0.000 | 0.000 | 0.000 | 0.000 |
| $E_{exc}$ (eV) | -0.094[a] | -0.027[a] | 0.032[a] | 0.271[a] ; 0.026[b] |
| $E_b$ (eV) | 0.094 | 0.027 | -- | -- |
| EI states | EI | EI | Non-EI | Non-EI |

[a] exciton momentum transfer from $\Gamma$ to M

[b] exciton with zero momentum transfer

Table 2. The calculated excitonic-insulator (EI) states of monolayer (1L), bilayer (2L), and trilayer (3L) 1T-$HfTe_2$ obtained using the $r^2$SCAN+SOC and mBSE methods with rescaled screening parameters α and μ. Here, $\varepsilon_{\infty\parallel}^{SC}$ and $\varepsilon_{\infty\perp}^{SC}$ denote the in-plane and out-of-plane components of the ion-clamped, long-wavelength dielectric constants of the supercell, respectively. $\varepsilon_{\infty\parallel}^{2D}$ and $\varepsilon_{\infty\perp}^{2D}$ are the corresponding rescaled two-dimensional values, and $\varepsilon_{\infty}^{2D,eff}$ is the effective 2D dielectric constant. $E_{exc}$ is the calculated exciton energy corresponding to the lowest-energy exciton with momentum transfer from Γ to M. $t$ is the thickness of the layered system, $c$ is the thickness of the supercell, $a$ is the optimized parameter used in the calculation, and $N_e$ is the number of valence electrons in the layered system.

Units: $t$(Å), $c$(Å), μ (Å$^{-1}$), $E_{exc}$(eV).

| | t | c | $a$ | $N_e$ | $\varepsilon_{\infty\parallel}^{SC}$ | $\varepsilon_{\infty\perp}^{SC}$ | $\varepsilon_{\infty\parallel}^{2D}$ | $\varepsilon_{\infty\perp}^{2D}$ | $\varepsilon_{\infty}^{2D,eff}$ | $\alpha$ | $\mu$ | $E_{exc}$ | EI status |
|---|---|---|---|---|---|---|---|---|---|---|---|---|---|
| 1L | 6.779 | 35 | 0.3 | 4 | 3.889 | 1.168 | 15.914 | 3.896 | 6.260 | 0.160 | 0.520 | -0.018 | EI |
| 1L | 1.1×6.779 | 35 | 0.3 | 4 | 3.889 | 1.168 | 14.558 | 3.084 | 5.090 | 0.196 | 0.531 | -0.037 | EI |
| 2L | 2×6.779 | 40 | 0.3 | 8 | 6.420 | 1.369 | 16.989 | 4.878 | 7.580 | 0.132 | 0.644 | 0.024 | Non EI |
| 2L | 2.25×6.779 | 40 | 0.3 | 8 | 6.420 | 1.369 | 15.213 | 3.409 | 5.570 | 0.180 | 0.662 | -0.0023 | EI |
| 3L | 3.1×6.779 | 43 | 0.3 | 12 | 8.974 | 1.649 | 17.314 | 5.129 | 7.914 | 0.126 | 0.735 | 0.046 | Non EI |
| 3L | 3.4×6.779 | 43 | 0.3 | 12 | 8.974 | 1.649 | 15.875 | 3.759 | 6.079 | 0.164 | 0.751 | 0.025 | Non EI |

Figures

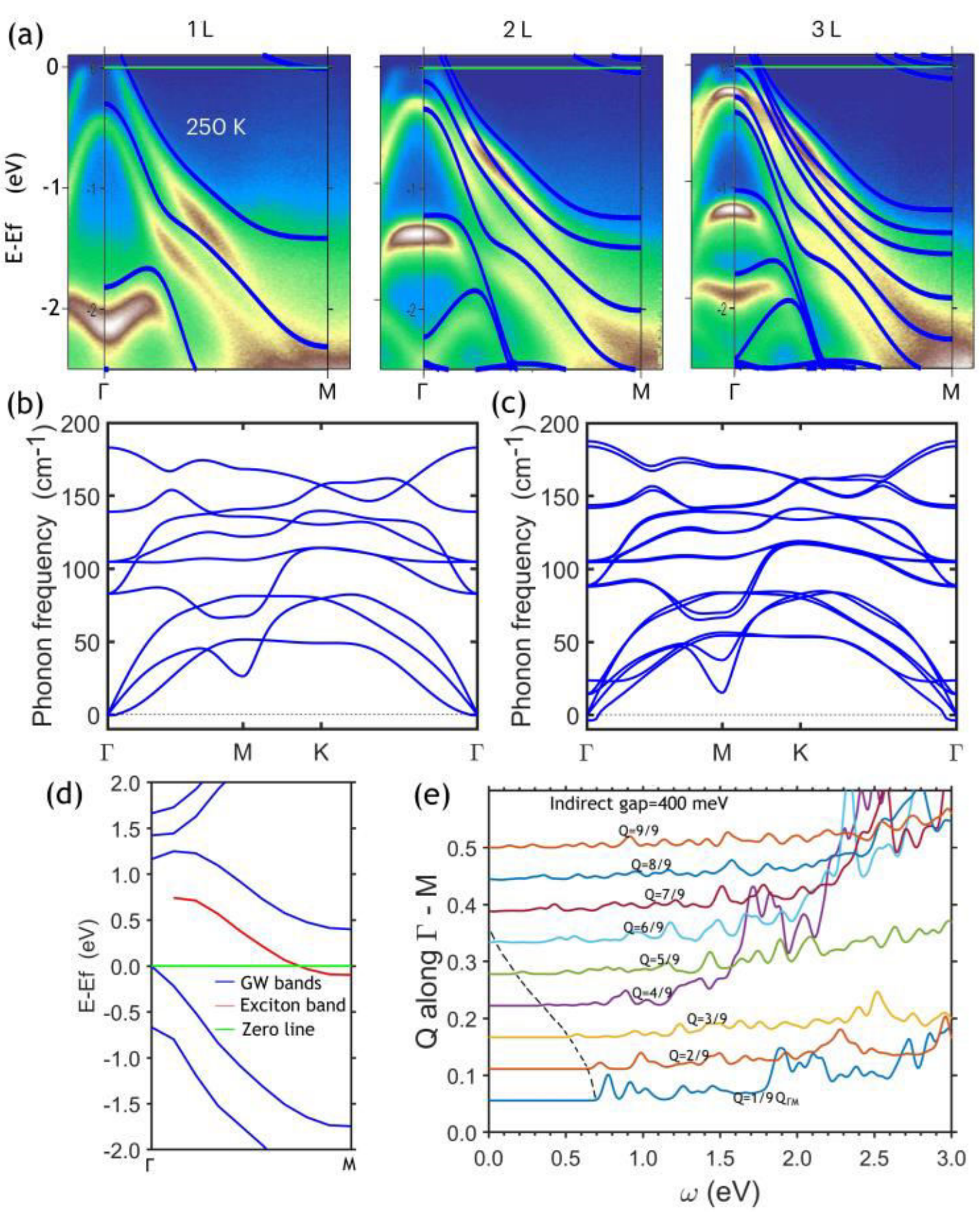

Figure 1. Multilayer 1T-$HfTe_2$ band structures, phonon dispersions, and EELS. (a) Calculated band structures (blue curves) along the Γ–M path superimposed on experimental ARPES spectra for monolayer (left), bilayer (middle), and trilayer (right) 1T-$HfTe_2$. (b,c) Phonon dispersions for the monolayer and bilayer, respectively. (d) $G_0W_0$ band structure along Γ–M of monolayer 1T-$HfTe_2$, together with the exciton band corresponding to the lowest exciton energies. (e) Calculated electron energy-loss spectra (EELS) for monolayer 1T-$HfTe_2$ at different momentum transfers, showing soft-mode–like features (black dashed curve). Adjacent curves are shifted vertically by 0.5/9 for clarity. Here, $Q = 5/9$ denotes a momentum transfer of $Q = 5\,Q_{\Gamma M}/9$, and so on, where $Q_{\Gamma M}$represents the wavevector from Γ to M.

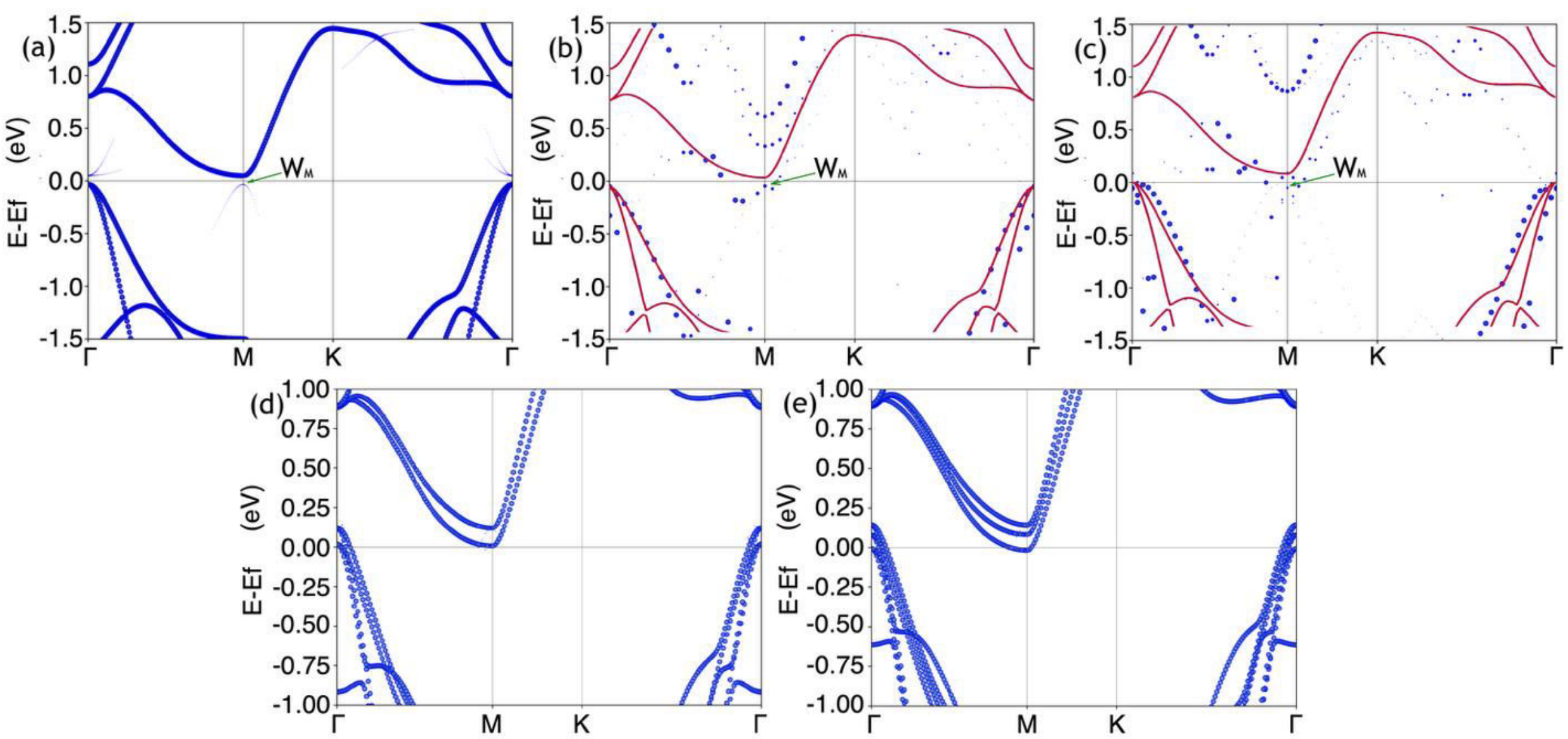

Figure 2. Unfolded band structures calculated using different approaches. All calculations are performed using 2×2×1 supercells and unfolded back to the 1×1×1 cell. (a) Structural symmetry breaking (SSB), in which atomic positions are perturbed from their symmetric locations. (b) Electronic symmetry breaking (ESB), where the atoms remain at their symmetric positions while electrons are promoted from the highest valence band to the lowest conduction band. (c) Combined structural and electronic symmetry breaking (SESB), in which both atomic position perturbations and electron promotions are applied. Panels (a)–(c) correspond to monolayer 1T-$HfTe_2$. The red curves in (b) and (c) show the band structure of monolayer 1T-$HfTe_2$ calculated using $r^2$SCAN and are included for comparison. (d, e) SSB calculations for bilayer and trilayer 1T-$HfTe_2$, respectively, obtained using the same procedure as in (a).

# Elusive Exciton Insulator States in 1T-$HfTe_2$: Exciton softening, and Symmetry Breaking by *Ab Initio* Methods

Hong Tang[1], Niraj Pangeni[1], Daniel D. Rivera[2], Adrienn Ruzsinszky[1]

[1]Department of Physics and Engineering Physics, Tulane University, New Orleans, LA 70118

[2]Instituto de Física, Universidade de São Paulo, 05315-970 São Paulo, São Paulo, Brasil

# Supplemental Information

**Note: Computational details**

Calculations for multilayer and bulk 1T $HfTe_2$ are performed using the Vienna *ab initio* Simulation Package (VASP) [1] with projector augmented-wave (PAW) pseudopotentials [2]. For the transition-metal Hf atoms, the *s* and *p* semicore states are included in the valence. The $r^2$SCAN meta-GGA functional [3] is employed throughout. For multilayer calculations, a vacuum spacing of 28 Å is applied along the out-of-plane direction to avoid spurious image interactions.

For phonon calculations of multilayers, a 40×40×1 k-point mesh, a plane-wave energy cutoff of 550 eV, and a force convergence criterion of 0.001 eV $Å^{-1}$ are used for structural relaxation of the primitive cell. Supercells of size 4×4×1 are employed for finite-displacement calculations, and the force constants and phonon modes are obtained using the Phonopy code [4]. During structural relaxations and phonon calculations, van der Waals (vdW) interactions are included using the rVV10 scheme [5,6] for both multilayer and bulk systems.

A 24×24×1 k-point mesh for multilayers and a 25×25×13 mesh for bulk are used for electronic self-consistent ground-state and band-structure calculations. Spin–orbit coupling (SOC) is included in the band-structure calculations. The calculated band structures, orbital-resolved band structures, and densities of states (DOS) for monolayer, bilayer, trilayer, and bulk 1T $HfTe_2$ are shown in Figs. S1–S4. The calculated phonon dispersions for monolayer and bilayer systems are shown in Fig. 1 of the main text, while the phonon dispersion of the trilayer is presented in Fig. S5.

Meta-GGA-based mBSE calculations [7–9] for excitonic properties are also performed using VASP. The ground-state wavefunctions are obtained with $r^2$SCAN+SOC. Dielectric screening is modeled analytically using two parameters, $\alpha$ and $\mu$, which are determined by fitting the wavevector-dependent dielectric function calculated from the response function at the random-phase approximation (RPA) level. The fitting curves for monolayer, bilayer, trilayer, and bulk systems are shown in Figs. S6–S9, and the corresponding fitted values of $\alpha$ and $\mu$ are summarized in Table S1.

For the structural symmetry-breaking (SSB) calculations, a 2×2×1 supercell of monolayer 1T $HfTe_2$ is constructed from the $r^2$SCAN-relaxed primitive 1×1×1 cell. All atomic positions in the 2×2×1 supercell are then randomized with a maximum displacement of 0.03 Å using the VASPKIT code [10]. Subsequently, the atomic positions in the randomized supercell are fully relaxed using $r^2$SCAN, while keeping the supercell lattice vectors fixed.

With appropriate k-point meshes generated by VASPKIT, self-consistent $r^2$SCAN calculations are performed, and the resulting band structures are back-folded to the primitive 1×1×1 Brillouin zone using VASPKIT. An energy cutoff of 550 eV and a nonzero-weight k-point mesh of 24×24×1 are used in all calculations.

In addition to the 2×2×1 supercell, SSB calculations are also carried out using a 3×3×1 supercell, following the same procedure. The corresponding results are shown in Fig. S10, where no unfolded band features are observed.

SSB calculations are further performed for bilayer and trilayer 1T $HfTe_2$. Table S2 summarizes the total energy differences of the 2×2×1 supercells between the SSB case and the symmetric (no-SSB) case for monolayer (1L), bilayer (2L), and trilayer (3L) 1T $HfTe_2$. The no-SSB calculations follow the same procedure as the SSB calculations, except that no atomic randomization is applied prior to relaxation. For the monolayer, the SSB configuration exhibits a slightly lower total energy than its corresponding no-SSB counterpart. For the bilayer and trilayer, the SSB configurations exhibit slightly higher total energies than their corresponding no-SSB counterparts. In all cases, the energy difference is very small ($< 0.002$ meV/atom).

For the electronic symmetry-breaking (ESB) calculations, the $2 \times 2 \times 1$ supercell is constructed from the $r^2$SCAN-relaxed primitive $1 \times 1 \times 1$cell, without any atomic-position randomization. Electrons from the highest valence band are promoted to the lowest conduction band to create electron–hole pairs. A hybrid r2SCAN-Y functional (where Y denotes the fraction of exact Hartree–Fock exchange, $Y = 0.286$; see α in Table S1) is used in the self-consistent calculations of the $2 \times 2 \times 1$ supercell, while all other computational procedures are the same as those used in the SSB calculations.

It is found that in the ESB calculations performed with $r^2$SCAN (without the hybrid $r^2$SCAN-Y functional), no unfolded band features appear around the Γ and M points (see Fig. S11). In this case, the band gap between Γ and M increases to 0.23 eV, compared to 0.082 eV obtained with $r^2$SCAN without electron promotion from the valence to the conduction bands.

For the combined structural and electronic symmetry-breaking (SESB) calculations, the relaxed 2×2×1 supercell obtained from the SSB approach is used. Both perturbed atomic positions and electron promotions are included, and the hybrid $r^2$SCAN-Y functional is employed.

$G_0W_0$ [11] and $G_0W_0$ + BSE [12] calculations for monolayer 1T $HfTe_2$ are performed using the BerkeleyGW package [11], interfaced with Quantum ESPRESSO [13]. The starting mean-field (MF) wavefunctions for the GW calculations are obtained using the PBE (Perdew–Burke–Ernzerhof) functional [14], including spin–orbit coupling and a Hubbard U correction (U = 3.2 eV). An energy cutoff of 78 Ry (~1061 eV) and an 18×18×1 k-point mesh are used for the MF wavefunction calculations. A cutoff energy of 16 Ry is adopted for dielectric screening.

A total of 440 bands is included in the band summation, and the exact static remainder correction [15] is employed to accelerate convergence with respect to the number of unoccupied bands. A vacuum spacing of 28 Å is applied along the out-of-plane direction, together with Coulomb truncation [11], to minimize spurious image interactions. For the optical calculations, four valence bands and four conduction bands are included.

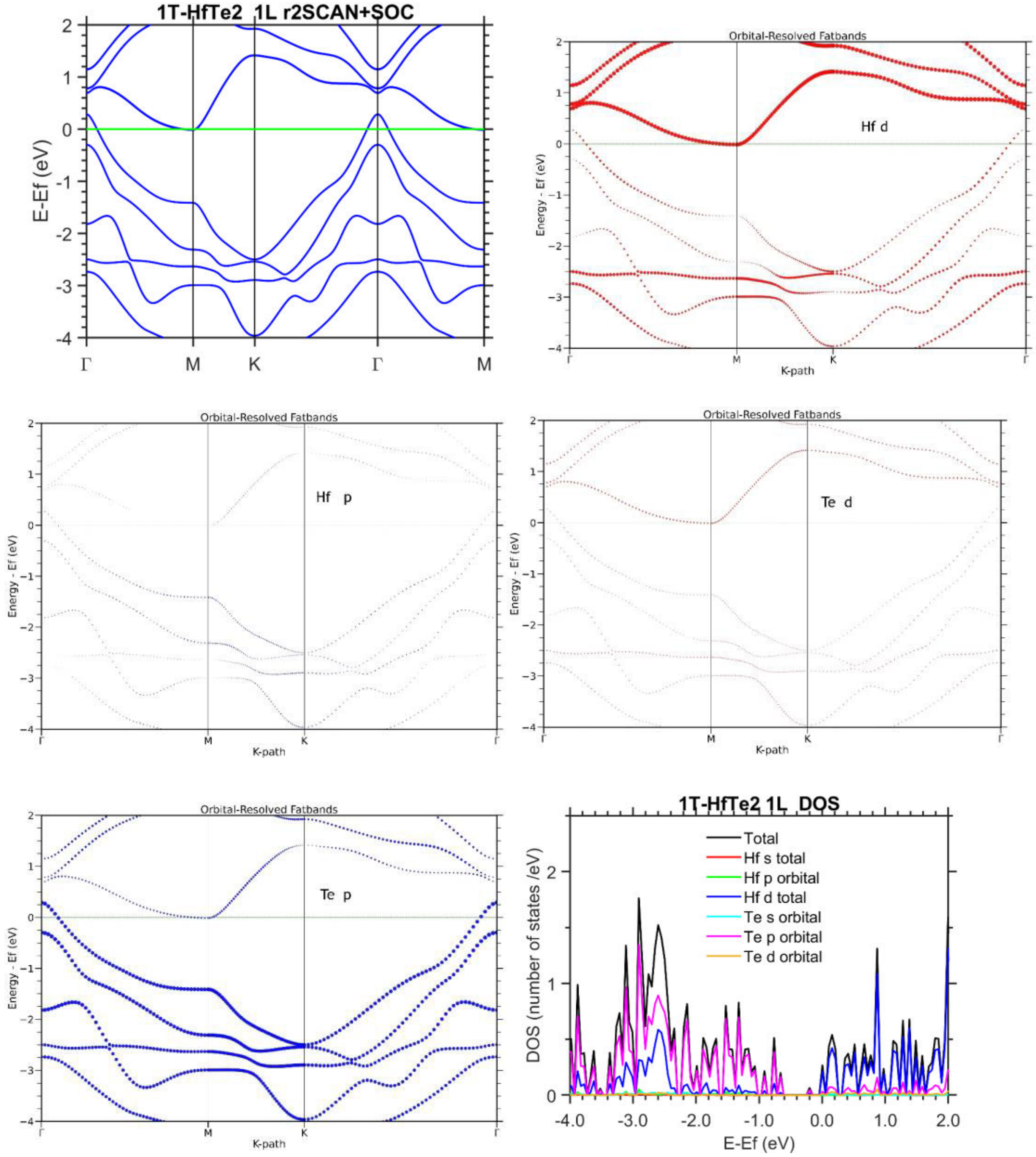

Figure S1. Calculated band structures and density of states (DOS) of monolayer 1T-$HfTe_2$ using $r^2$SCAN+SOC. The top left panel shows the conventional band structure. The top right, middle left, middle right, and bottom left panels show atomic-orbital-resolved band structures, with the marker size proportional to the magnitude of the orbital projection. The bottom right panel shows the DOS with contributions from different atomic orbitals.

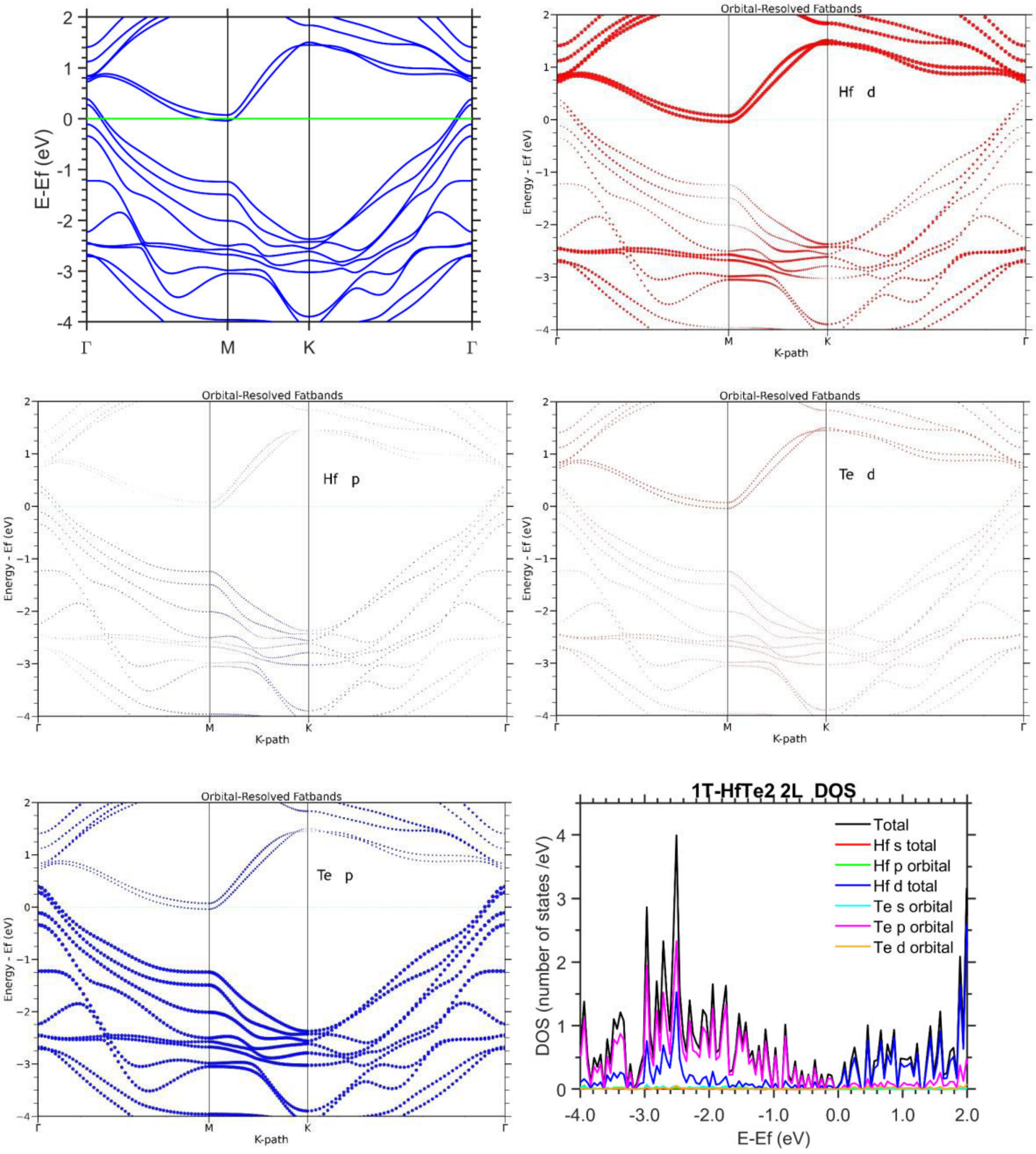

Figure S2. Calculated band structures and density of states (DOS) of bilayer 1T-$HfTe_2$ with $r^2$SCAN+SOC. The panels are arranged in the same way as in Figure S1.

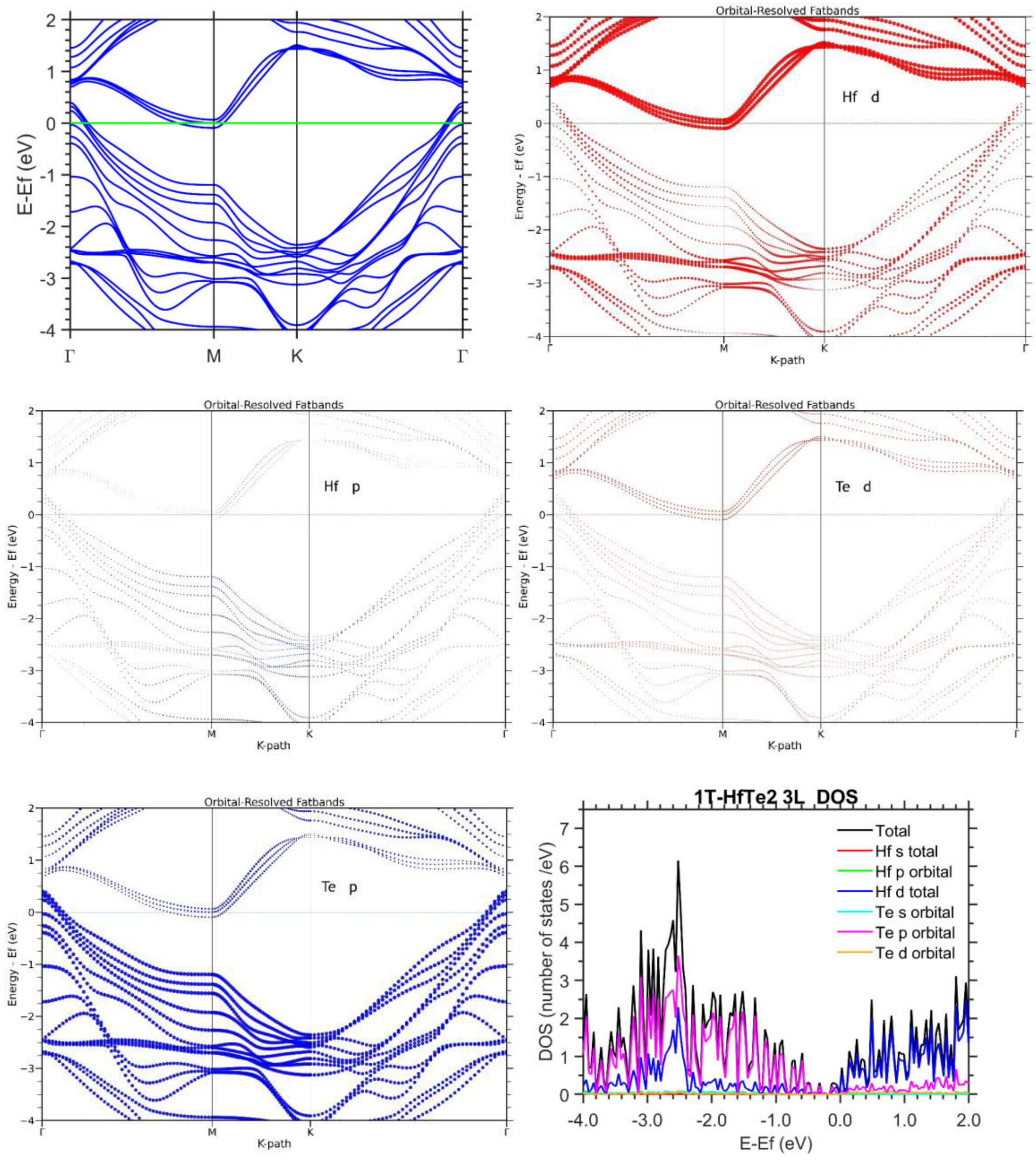

Figure S3. Calculated band structures and density of states (DOS) of tri-layer 1T-$HfTe_2$ with $r^2$SCAN+SOC. The panels are arranged in the same way as in Figure S1.

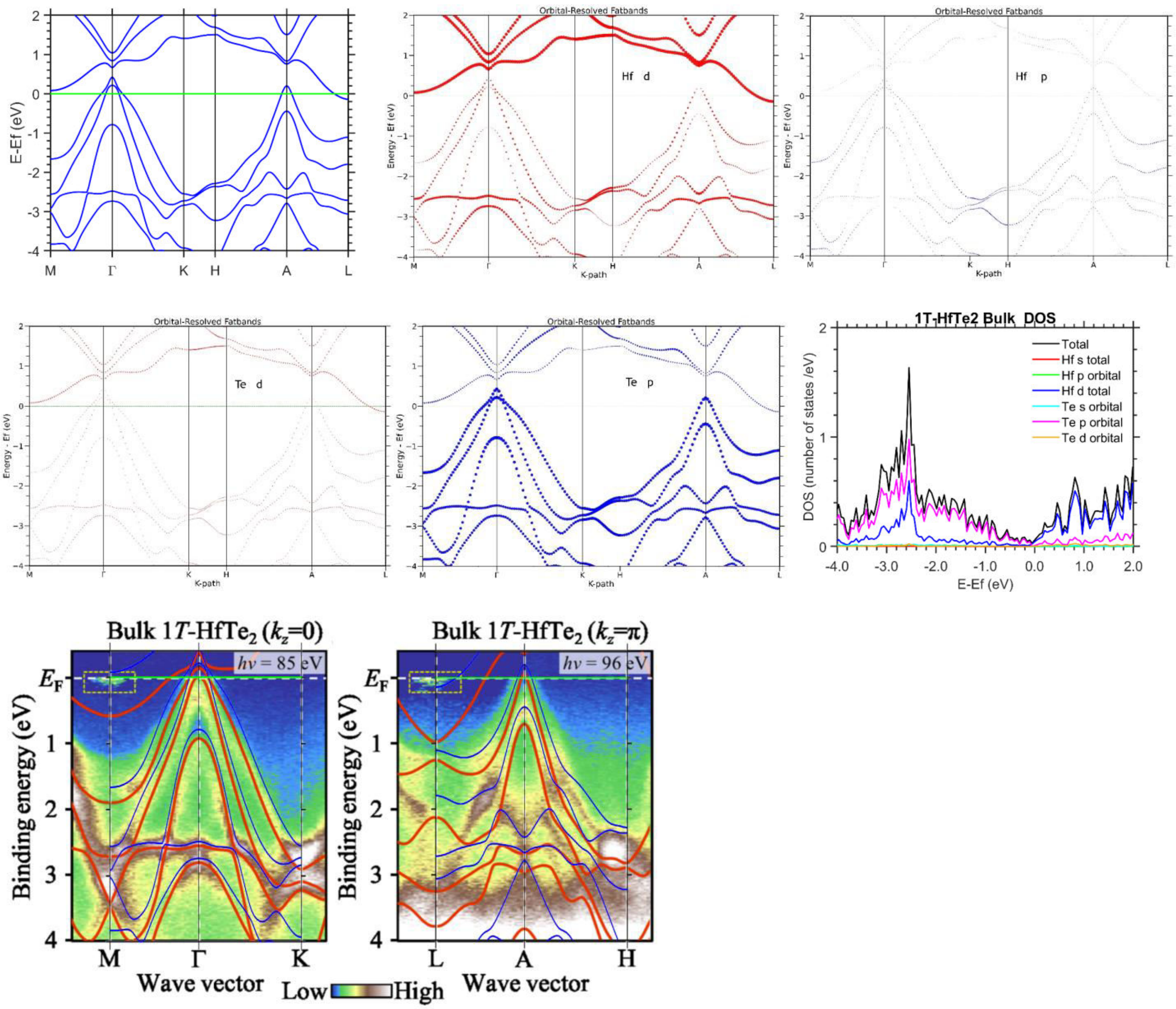

Figure S4. Calculated band structures and densities of states (DOS) of bulk 1T $HfTe_2$ using r²SCAN+SOC. The bottom row shows the calculated band structures superimposed on experimental ARPES images, where the blue curves correspond to r²SCAN+SOC and the red curves to PBE+SOC. The ARPES and PBE+SOC data are taken from Ref. [16].

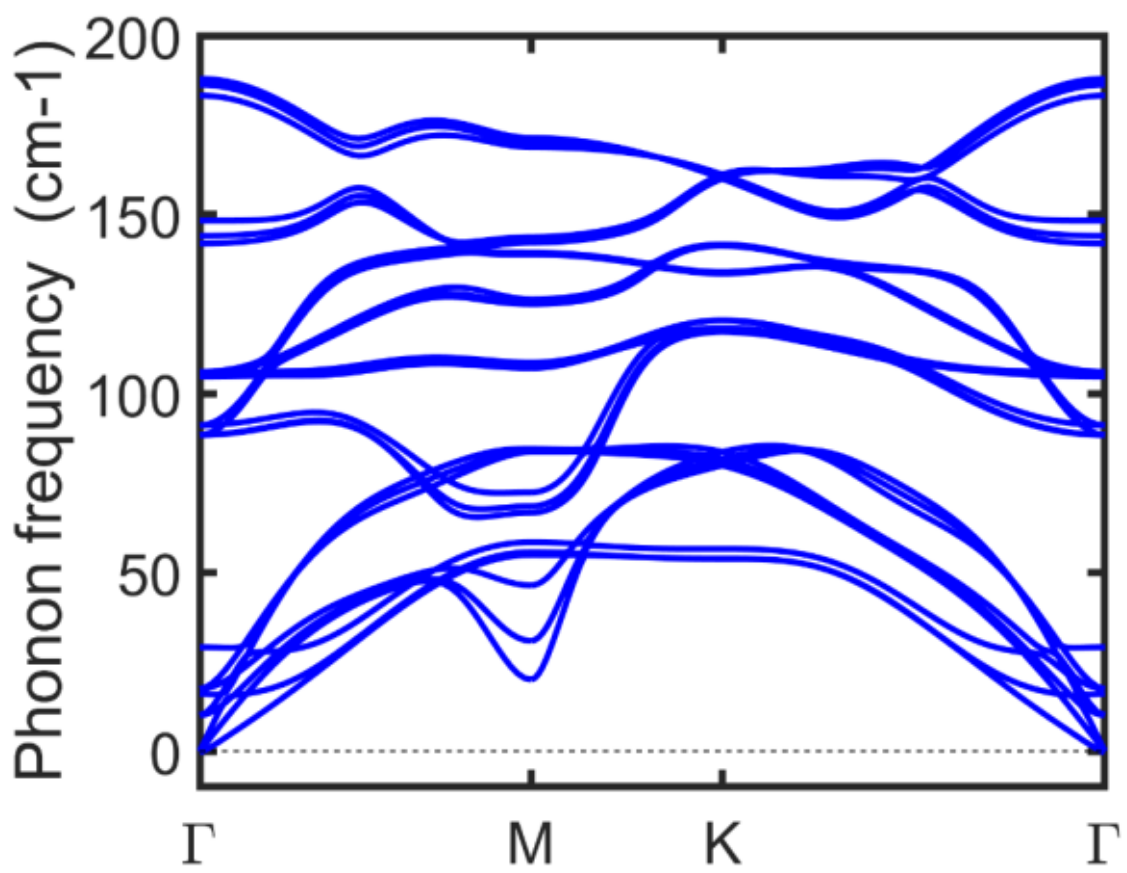

Figure S5. Calculated phonon dispersion of tri-layer 1T-$HfTe_2$ with $r^2$SCAN+rVV10.

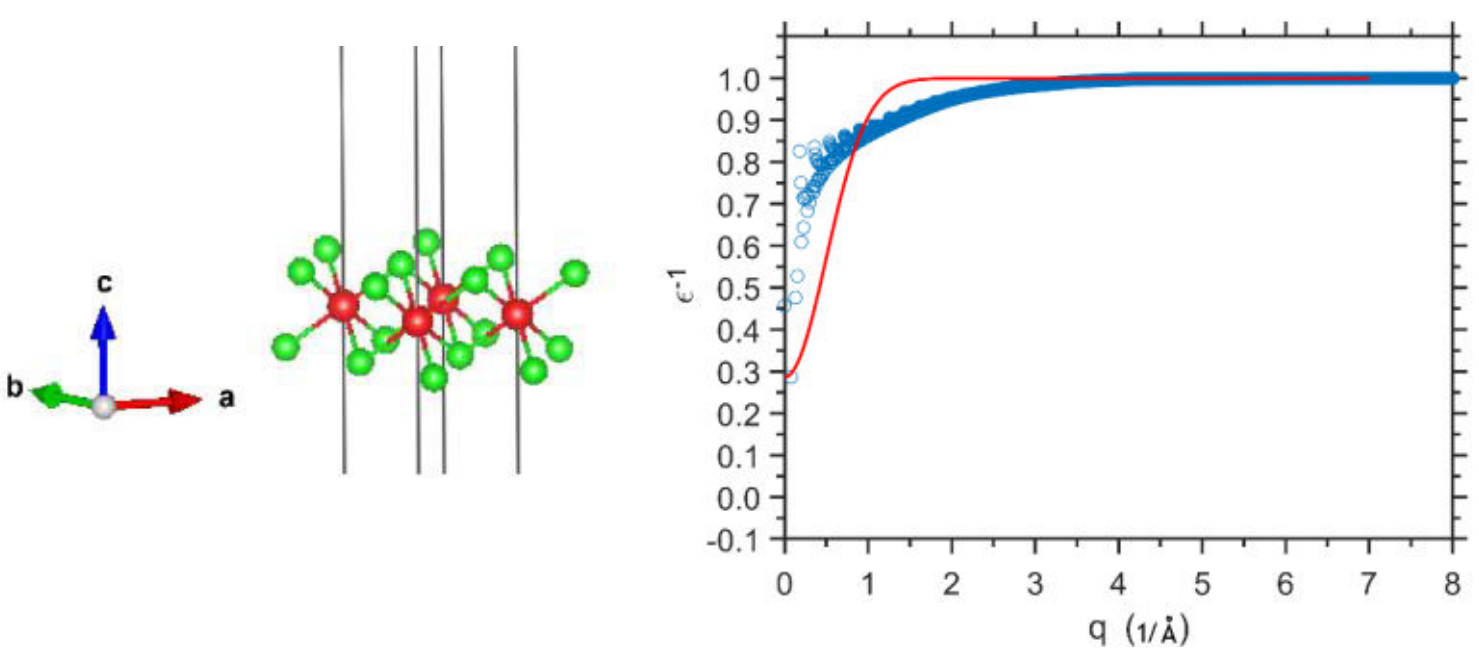

Figure S6. Left: Atomic structure of monolayer 1T $HfTe_2$, with Hf atoms shown in red and Te atoms in green. Right: Fit (red curve) of the dielectric function model $\varepsilon^{-1}(q) = 1 - (1 - \alpha)\exp(-q^2/(4\mu^2))$ to the calculated dielectric function obtained with $r^2$SCAN+SOC (blue circles).

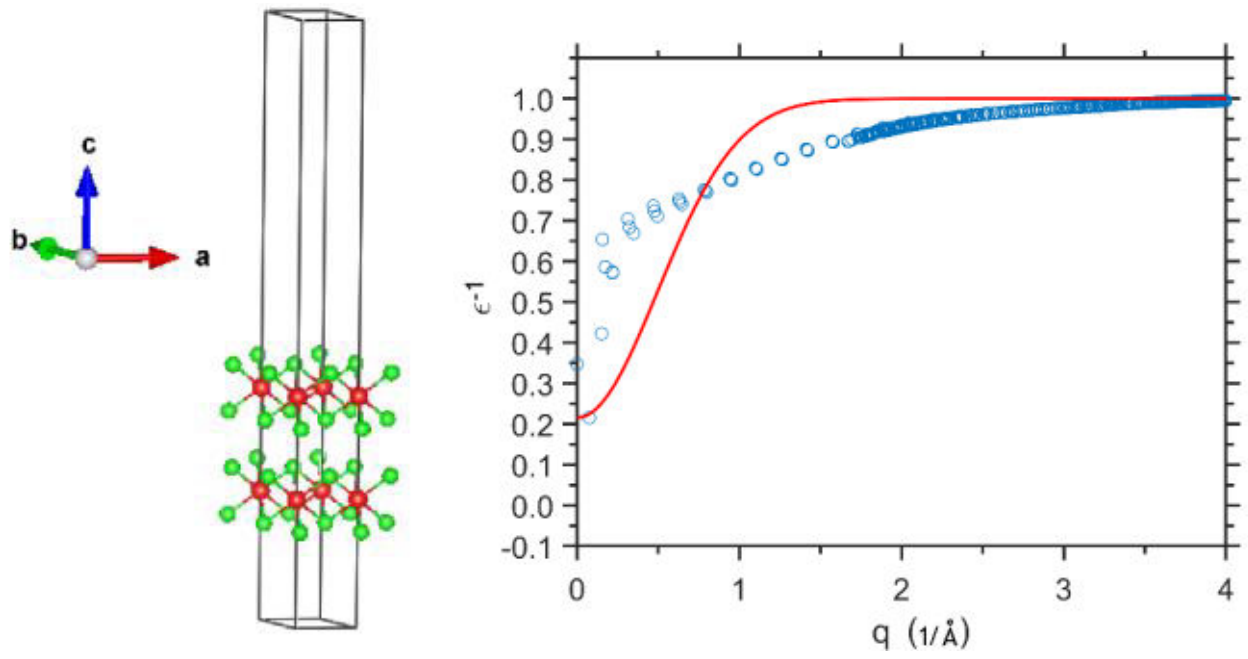

Figure S7. For bilayer 1T-HfTe2, arranged similarly to Figure S6.

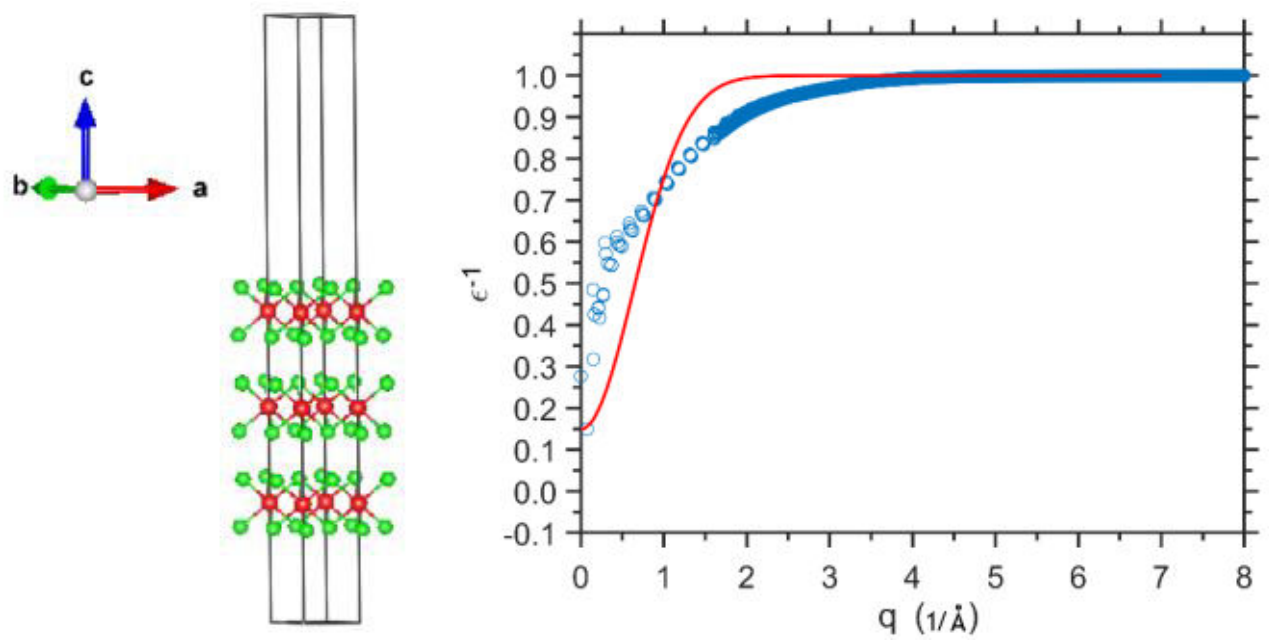

Figure S8. For tri-layer 1T-HfTe2, arranged similarly to Figure S6.

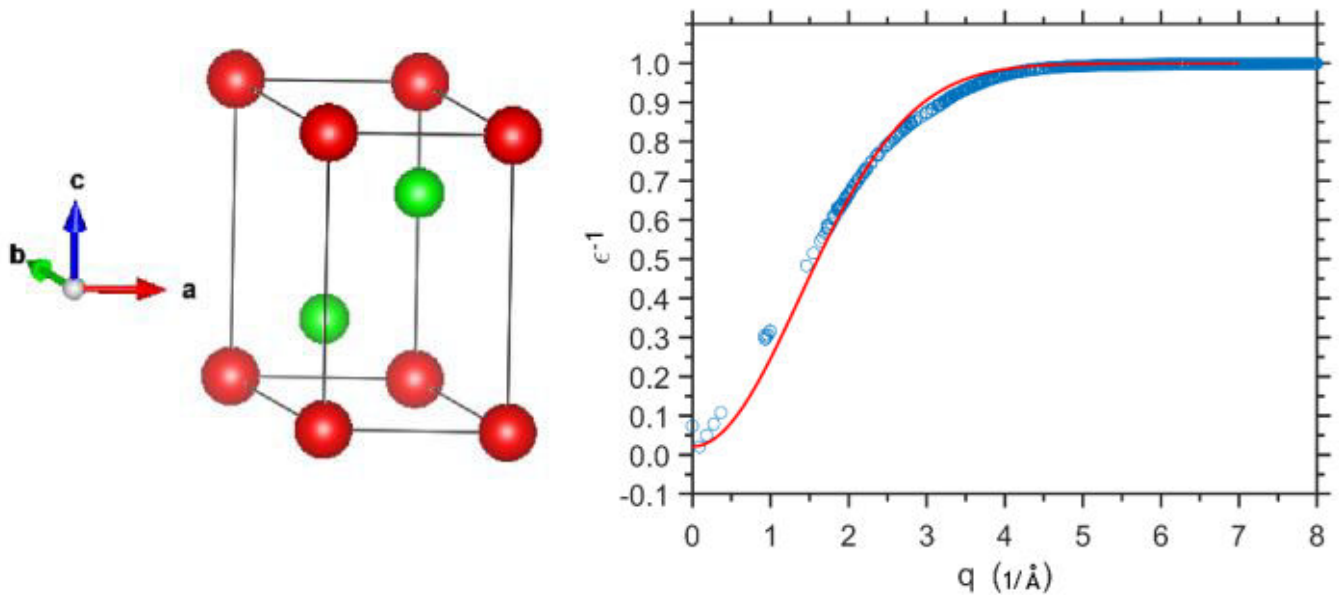

Figure S9. For bulk 1T-HfTe2, arranged similarly to Figure S6.

Table S1. The fitted parameters $\alpha$ and $\mu$, obtained by fitting the dielectric function model $\varepsilon^{-1}(q) = 1 - (1-\alpha)\exp\left(-\frac{q^2}{4\mu^2}\right)$ for the 1T-HfTe2 multilayers and bulk.

| | Monolayer 1T-HfTe2 | Bilayer 1T-HfTe2 | Tri-layer 1T-HfTe2 | Bulk 1T-HfTe2 |
|---|---|---|---|---|
| $\alpha$ | 0.286 | 0.215 | 0.149 | 0.021 |
| $\mu$ (1/Å) | 0.35 | 0.35 | 0.45 | 0.98 |

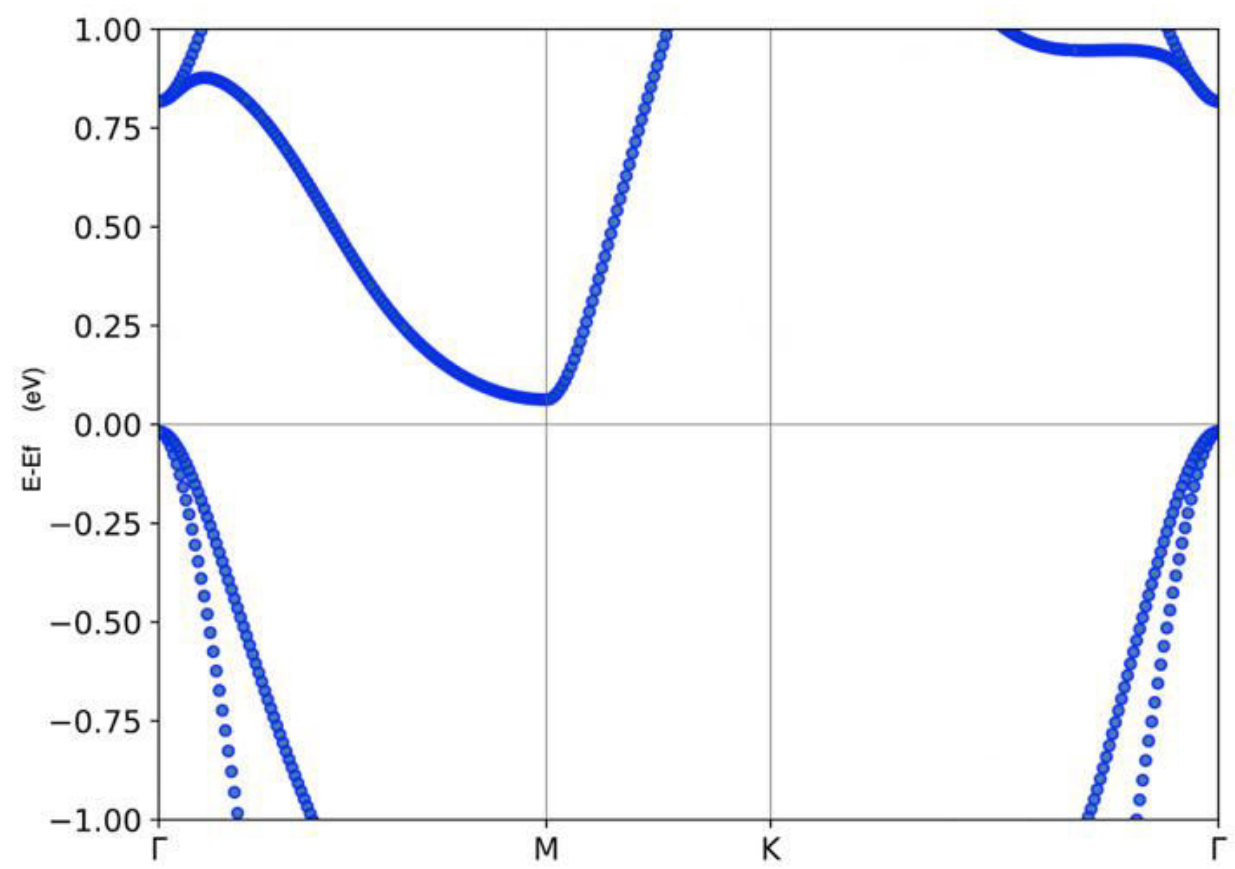

Figure S10. Structural symmetry-breaking (SSB) approach applied to a 3×3×1 supercell of monolayer 1T $HfTe_2$. The plot shows the back-folded band structure in the 1×1×1 cell obtained from a randomized and relaxed 3×3×1 supercell. No unfolded band features are observed around the Γ and M points.

Table S2. The total energy difference (ΔE) of the 2×2×1 supercell between the structural symmetry-breaking (SSB) case and the symmetric (no-SSB) case for monolayer (1L), bilayer (2L), and trilayer (3L) 1T $HfTe_2$. All results are obtained using 2×2×1 supercells.

| | 1L | 2L | 3L |
|---|---|---|---|
| SSB (eV) | -406.62154077 | -807.62619123 | -1212.00046674 |
| Symmetric (eV) | -406.62153615 | -807.62621900 | -1212.00050096 |
| $\Delta E$ (eV/cell) | -0.00000462 | 0.00002777 | 0.00003422 |
| $\Delta E$ (eV/atom) | -0.00000039 | 0.00000116 | 0.00000095 |

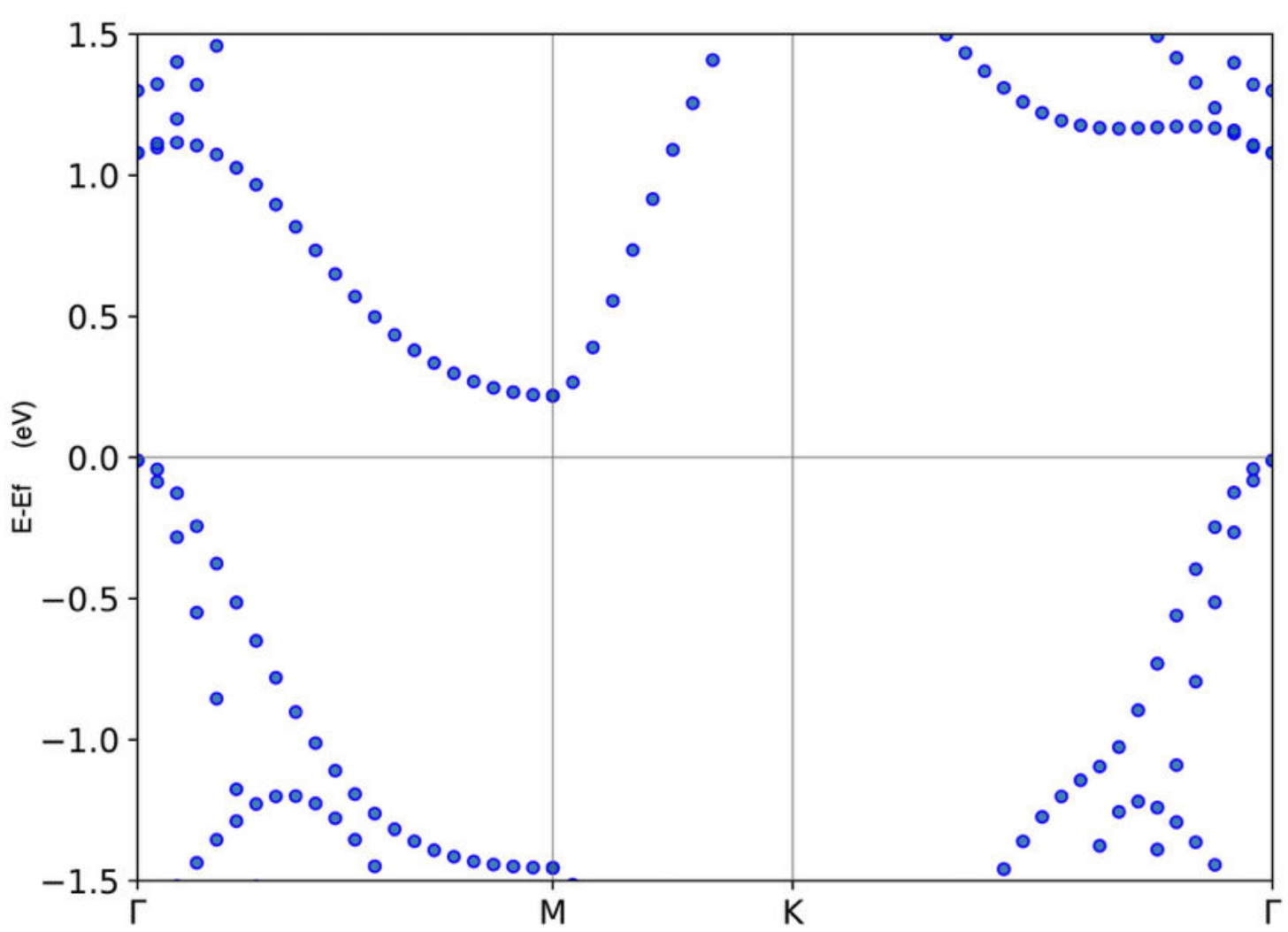

Figure S11. The electronic symmetry-breaking (ESB) approach applied to a symmetric 2×2×1 supercell of monolayer 1T $HfTe_2$ using r²SCAN. No hybrid r²SCAN-Y functional is used. In the symmetric 2×2×1 supercell, electrons from the entire highest valence band are promoted to the lowest conduction band, and no atomic randomization is performed. No unfolded band features are observed around the Γ and M points. A gap of 0.23 eV opens between Γ and M.

References

1. Kresse, G.; Furthmüller, J. Efficient Iterative Schemes for Ab Initio Total-Energy Calculations Using a Plane-Wave Basis Set, Phys. Rev. B **1996,** 54, 11169.
2. Kresse, G.; Joubert, D. From Ultrasoft Pseudopotentials to the Projector Augmented-Wave Method, Phys. Rev. B **1999**, 59, 1758.
3. Furness, J. W.; Kaplan, A. D.; Ning, J.; Perdew, J. P.; Sun, J. Accurate and Numerically Efficient r2SCAN Meta-Generalized Gradient Approximation, J. Phys. Chem. Lett. **2020**, 11, 8208−8215.
4. Atsushi Togo, Laurent Chaput, Terumasa Tadano, and Isao Tanaka, Implementation strategies in phonopy and phono3py, J. Phys. Condens. Matter 35, 353001-1-22 (2023).
5. R. Sabatini, T. Gorni, and S. de Gironcoli, Nonlocal van der Waals density functional made simple and efficient, Phys. Rev. B 87, 041108(R) (2013).
6. J. Ning, M. Kothakonda, J. W. Furness, A. D. Kaplan, S. Ehlert, J. G. Brandenburg, J. P. Perdew, and J. Sun, Workhorse minimally empirical dispersion-corrected density functional with tests for weakly bound systems: r²SCAN+rVV10, Phys. Rev. B 106, 075422 (2022).
7. A. Tal, P. Liu, G. Kresse, and A. Pasquarello, Accurate optical spectra through time-dependent density functional theory based on screening-dependent hybrid functionals, Phys. Rev. Research 2, 032019(R) (2020).

8. W. Chen, G. Miceli, G.M. Rignanese, and A. Pasquarello, Nonempirical dielectric-dependent hybrid functional with range separation for semiconductors and insulators, Phys. Rev. Mater. 2, 073803 (2018).
9. Tang, H.; Yin, L.; Csonka, G. I.; Ruzsinszky, A. Exploring the exciton insulator state in 1T-TiSe2 monolayer with advanced electronic structure methods, Phys. Rev. B 2025, 111, L201401.
10. V. Wang, N. Xu, J.-C. Liu, G. Tang, W.-T. Geng, VASPKIT: A User-Friendly Interface Facilitating High-Throughput Computing and Analysis Using VASP Code, Computer Physics Communications 267, 108033 (2021).
11. Deslippe, J.; Samsonidze, G.; Strubbe, D. A.; Jain, M.; Cohen, M. L.; Louie, S. G. BerkeleyGW: A Massively Parallel Computer Package for the Calculation of the Quasiparticle and Optical Properties of Materials and Nanostructures, Comput. Phys. Commun. **2012**, 183, 1269.
12. Rohlfing, M.; Louie, S. G. Electron-Hole Excitations and Optical Spectra from First Principles, Phys. Rev. B, **2000**, 62, 4927.
13. Giannozzi, P.; Andreussi, O.; Brumme, T.; Bunau, O.; Buongiorno Nardelli, M.; Calandra, M.; Car, R.; Cavazzoni, C.; Ceresoli, D.; Cococcioni, M.; Colonna, N.; Carnimeo, I.; Dal Corso, A.; de Gironcoli, S.; Delugas, P.; DiStasio Jr, R. A.; Ferretti, A.; Floris, A.; Fratesi, G.; Fugallo, G.; Gebauer, R. et al. Advanced Capabilities for Materials Modelling with Quantum ESPRESSO, J. Phys.: Condens. Matter. **2017**, 29, 465901.
14. Perdew, J. P.; Burke, K.; Ernzerhof, M. Generalized Gradient Approximation Made Simple, Phys. Rev. Lett. **1996**, 77, 3865.
15. Deslippe, J.; Samsonidze, G.; Jain, M.; Cohen, M. L.; Louie, S. G. Coulomb-hole summations and energies for $GW$ calculations with limited number of empty orbitals: A modified static remainder approach, Phys. Rev. B **2013**, 87, 165124.
16. Y. Nakata, K. Sugawara, A. Chainani, K. Yamauchi, K. Nakayama, S. Souma, P.-Y. Chuang, C.-M. Cheng, T. Oguchi, K. Ueno, T. Takahashi, and T. Sato, Dimensionality reduction and band quantization induced by potassium intercalation in 1T-HfTe2, Phys. Rev. Materials 3, 071001(R) (2019).